\documentstyle[sprocl]{article}

\input{psfig}

\def \ms {{\overline{\mbox{MS}}}}

\def\be{\begin{equation}}
\def\ee{\end{equation}}
\def\bea{\begin{eqnarray}}
\def\eea{\end{eqnarray}}

\def \ms {{\overline{\mbox{MS}}}}

\begin{document}

\begin{flushright} {
\bf US-FT / 06-01 \\ 

} \end{flushright}

\title{HIGHER TWIST OPERATOR EFFECTS TO PARTON DENSITIES AT SMALL $X$}

%

\author{A. V. KOTIKOV}

\address{Bogoliubov Lab. of Theor. Phys., JINR \\
141980 Dubna, Russia,~
E-mail: kotikov@thsun1.jinr.de}

\author{G. PARENTE}

\address{Dep.
de F\'\i sica de Part\'\i culas, 
Univ.
de Santiago de Compostela\\
15706 Santiago de Compostela, Spain,~
E-mail: gonzalo@fpaxp1.usc.es}


\maketitle\abstracts{ 
We investigate  the  $Q^2$ evolution of parton distributions
at small $x$ values,
obtained in the case
of soft initial conditions.
The contributions of twist-two and (renormalon-type) higher-twist 
operators of the Wilson operator product expansion are taken into account.
The results are in very good agreement with deep inelastic scattering
experimental data from HERA.}

\vskip -0.5cm
The measurements of the deep-inelastic scattering
structure function
(SF)
$F_2$ in HERA
\cite{H1}
have permitted the access to
a very interesting kinematical range for testing the theoretical
ideas on the behavior of quarks and gluons carrying a very low fraction
of momentum of the proton, the so-called small $x$ region.
The reasonable agreement between HERA data and the 
next-to-leading order (NLO)
approximation of
perturbative
QCD that has been observed for $Q^2 \geq 2 $GeV$^2$ (see 
\cite{Q2evo} and references therein) indicates that
perturbative QCD could describe the SF
evolution 
up to very low $Q^2$ values.

The standard program \cite{MRS,KKPS} to study the 
behavior of
quarks and gluons
is carried out by comparison of data
with the numerical solution of the
DGLAP
equations 
by fitting the parameters of the
$x$ profile of partons at some initial $Q_0^2$. 
However, if one is interested in analyzing exclusively the
small $x$ region ($x \leq 0.01$), 
there is the alternative of doing a simpler analysis
by using some of the existing analytical solutions of DGLAP 
in the small $x$ limit (see, for example, \cite{BF1,Q2evo}).
The main ingredients of the study \cite{Q2evo} are:
\begin{itemize}
\item 
%
Both, the gluon and quark singlet densities are
presented in terms of two components ($'+'$ and $'-'$)
which are obtained from the analytical $Q^2$
dependent expressions of the corresponding ($'+'$ and $'-'$)
parton distributions moments.
\item
%
\vskip -0.3cm
The $'-'$ component is constant
at small $x$, whereas the 
$'+'$ component grows at $Q^2 \geq Q^2_0$ as 
\vskip -0.5cm
$$ 
\hspace*{-0.7cm}
\sim \exp{\left(2\sqrt{\left[
a_+\ln \left(
\frac{a_s(Q^2_0)}{a_s(Q^2)} \right) -
\left( b_+ +  a_+ \frac{\beta_1}{\beta_0} \right)
\Bigl( a_s(Q^2_0) - a_s(Q^2) \Bigr) \right] 
\ln \left( \frac{1}{x}  \right)} \right)},
$$
where the LO term $a_+ = 12/\beta_0$ and the NLO one $b_+ = 412f/(27\beta_0)$. 
Here the coupling constant
$a_s=\alpha_s/(4\pi)$, $\beta_0$ and $\beta_1$ are the first two 
coefficients of QCD 
$\beta$-function and $f$ is the number of active flavors.
\end{itemize}

We shortly compile below the main results 
of \cite{Q2evo} 
at the leading order (LO)
approximation 
and demonstrate some new (preliminary) results,
where
the contributions of higher-twist (HT) operators 
(i.e. twist-four ones and twist-six ones)
of
the Wilson operator product expansion are taken into account
in the framework of renormalon model (see \cite{Beneke}).
The importance of the HT contributions 
at small-$x$ has been demonstrated in 
\cite{Bartels}.
\vskip 0.3cm

{\bf 1.} Our purpose
is to show the small $x$ asymptotic
form of parton distributions
in the framework of the DGLAP equation starting at some $Q^2_0$ with
the flat function:
\vspace{-0.2cm}
 \begin{eqnarray}
f^{\tau2}_a (Q^2_0) ~=~
A_a ~~~~(\mbox{ hereafter } a=q,g), \label{1}
 \end{eqnarray}
where $f^{\tau2}_a$ are the leading-twist (LT) parts of
parton (quark and gluon)
distributions (PD) multiplied by $x$
and $A_a$ are unknown parameters that have to be determined from data.
Through this work at small $x$ we neglect
the non-singlet quark component
\footnote{
We would like to note that new HERA data \cite{H1} show a rise
of $F_2$ structure function at low $Q^2$ values ($Q^2 \sim 1 $GeV$^2$)
when $x \to 0$ (see Fig.1, for example). The rise can be explained
in a natural way by incorporation  of 
HT terms in our
analysis (see Eqs.(\ref{r1})-(\ref{r7})).}.

The full small $x$ asymptotic results
for PD
 and SF $F_2$ 
at LO 
is:
\vspace{-0.1cm}
\bea
F_2(x,Q^2) ~=~
e \cdot f_q(x,Q^2),~~~~
 f_a(x,Q^2) ~=~
f^{+}_a(x,Q^2) + 
f^{-}_a(x,Q^2) \; , 
\label{r11}
\end{eqnarray}
where the $'+'$ and $'-'$ components $f^{\pm}_a(x,Q^2)$
are given by the sum
 \begin{eqnarray}
 f^{\pm}_a(x,Q^2) ~=~ f^{\tau2,\pm}_a(x,Q^2) + 
f^{h\tau,\pm}_a(x,Q^2) \;  
\label{r12}
\eea
of
the LT
parts $f^{\tau2,\pm}_a(x,Q^2)$ 
and the HT
parts $f^{h\tau,\pm}_a(x,Q^2)$, 
respectively.



The small $x$ asymptotic results for PD, $f^{\tau2,\pm}_a$ 
\vspace{-0.2cm}
\bea
f^{\tau2,+}_g(x,Q^2)&=& \left(A_g + \frac{4}{9} A_q \right)
\tilde I_0(\sigma) \; e^{-\overline d_{+}(1) s} \,+\,O(\rho) 
~~\;\; ,\label{8.0} \\
f^{\tau2,+}_q(x,Q^2)&=& \frac{f}{9}\left(A_g + \frac{4}{9} A_q \right) 
\rho \; \tilde I_1(\sigma) \;
e^{-\overline d_{+}(1) s} \,+\,O(\rho) \; , \label{8.01} 
\\
f^{\tau2,-}_q(x,Q^2) &=&
A_q e^{- d_{-}(1) s} \,+\,O(x),~
f^{\tau2,-}_g(x,Q^2)~= - \frac{4}{9} f^{\tau2,-}_q(x,Q^2) 
\; ,\label{8.02}
\eea
where
$\overline d_{+}(1) = 1+20f/(27\beta_0)$ and
$          d_{-}(1) = 16f/(27\beta_0)$
are the regular parts of $d_{+}$ and $d_{-}$
anomalous dimensions, respectively, in the limit $n\to1$ 
\footnote{From now on, for a quantity $k(n)$ we use the notation
$\hat k(n)$ for the singular part when $n\to1$ and
$\overline k(n)$ for the corresponding regular part. }. 
The function $\tilde I_{\nu}$ ($\nu=0,1$) 
coincides with 
the modified Bessel
function $I_{\nu}$ at $s \geq 0$
and 
the Bessel function $J_{\nu}$ at $s < 0$.
Using the 
calculations
\cite{SMaMaS,method}, we show the HT effect
in the renormalon case.
We present the results only for the terms proportional of some power
of $\ln{(1/x)}$ (full expressions can be found in the last papaer of 
\cite{Q2evo}),   
making the following subtitutions
in the corresponding LT
results presented in
Eqs.(\ref{8.0})-(\ref{8.02}):

\vspace{0.2cm}
$f^{\tau2,+}_a(x,Q^2)$ (see Eqs.(\ref{8.0}),(\ref{8.01})) 
$\to f^{h\tau,+}_a(x,Q^2)$
~~ by
\vspace{-0.2cm}
\bea 
A_a \biggl\{\tilde I_0(\sigma),~\rho \tilde I_1(\sigma) \biggl\} \,\to\, 
A_a 
\biggl\{
\frac{32f}{15\beta_0^2},~\frac{256f}{45\beta_0^2}  \biggl\}
 \, \biggl( 
\frac{\Lambda^2_{1,a}}{Q^2} -\frac{8}{7} \frac{\Lambda^4_{2,a}}{Q^4}
\biggr)
\,\frac{1}{\rho} \tilde I_1(\sigma)
\,+\, ... ,  
\label{r1}
\eea
where
$\Lambda^2_{1,a}$ and $\Lambda^4_{2,a}$ are magnitudes of twist-four and 
twist-six corrections.

$f^{\tau2,-}_g(x,Q^2)$ (see Eq.(\ref{8.02}))
 $ \to f^{h\tau,-}_g(x,Q^2)$
~~ by
\vspace{-0.2cm}
\bea 
A_q ~\to ~ A_q \cdot
\frac{32f}{15\beta_0^2} \, \biggl( 
\frac{\Lambda^2_{1,q}}{Q^2} -\frac{8}{7} \frac{\Lambda^4_{2,q}}{Q^4}
\biggr)
\, \ln \left(\frac{1}{x} \right) + ~...~.
\label{r6}
\eea

$f^{\tau2,-}_q(x,Q^2)$ (see Eq.(\ref{8.02})) $ \to f^{h\tau,-}_q(x,Q^2)$
~~ by
\vspace{-0.2cm}
\bea 
A_q &\to & 
\frac{128f}{45\beta_0^2} \biggl\{ A_q \cdot
\biggl(
\frac{\Lambda^2_{1,q}}{Q^2} \biggl[
\ln \left(\frac{Q^2}{x\Lambda^2_{1,q}} \right)
- \frac{209}{60} -  \frac{8f}{81}
\biggl]
~-~
\frac{8}{7} \frac{\Lambda^4_{2,q}}{Q^4} 
\biggl[\ln \left(\frac{Q^2}{x\Lambda^2_{2,q}} \right)
\nonumber \\
& &~- \frac{6517}{3150} -  \frac{8f}{81}
\biggl] \biggl)
- \frac{2f}{9}
A_g \cdot
\biggl( 
\frac{\Lambda^2_{1,g}}{Q^2} -\frac{8}{7} \frac{\Lambda^4_{2,g}}{Q^4}
\biggr) \biggr\} \,
\ln \left(\frac{1}{x} \right) + ~... ~.
\label{r7}
\eea



\begin{figure}[t]
\psfig{figure=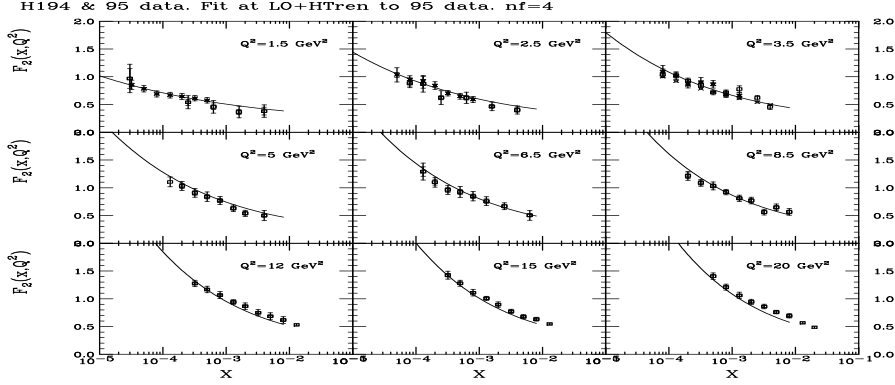,height=2.0in,width=4.7in}
\vskip -0.3cm
\caption{The structure function $F_2$ as a function of $x$ for different
$Q^2$ bins. The experimental points are from H1.
The inner error 
bars are statistic while the outer bars represent statistic and systimatic 
errors added in quadrature. 
The 
curves are obtained 
from fits at LO 
when the HT contributions 
have been incorporated.}
\vskip -0.5cm
\end{figure}

{\bf 2.}
With the help of the above equations
we have analyzed $F_2$ HERA data at small $x$ from the H1 collaboration.
We have fixed the number of active flavors $f$=4 and
$\Lambda_{\ms}(n_f=4) = 250$ MeV, which
is a reasonable value extracted from the traditional (higher $x$)
experiments.
Moreover, we put $\Lambda_{1,a}=\Lambda_{2,a}$ in agreement with
\cite{DaWe}.

The results are shown on Fig. 1. 
We found very good agreement between our approach based on QCD 
and HERA data.
The (renormalon-type) HT
terms lead to the natural explanation of
the rise of $F_2$ structure function at low values of $Q^2$ and $x$.

As a next step of our investigations, we plan to finish this study 
and to investigate HT contributions 
to PD and SF relations, 
observed
in \cite{KoPa,KOPAFL}.

\vskip 0.2cm
A.K. was supported by Alexander von Humboldt fellowship
and by DIS2001 Orgcommittee. 
G.P. was supported in part by Xunta de Galicia
(PGIDT00 PX20615PR) and CICYT (AEN99-0589-C02-02).

\vskip 0.2cm
{\bf References}

\end{document}